
\input harvmac

\def\CN{{\cal N}}
\def\CM{{\cal M}}
\def\IR{\relax{\rm I\kern-.18em R}}
\let\includefigures=\iftrue
%
%
%
\let\expandedversion=\iffalse
\includefigures
\message{If you do not have epsf.tex (to include figures),}
\message{change the option at the top of the tex file.}
\input epsf
\def\fig#1#2{\topinsert\epsffile{#1}\noindent{#2}\endinsert}
\def\ebox#1#2{\topinsert\epsfbox{#1}\noindent{#2}\endinsert}
\else
\def\fig#1#2{\vskip .5in
\centerline{Figure}
\vskip .5in}
\def\ebox#1#2{\vskip .5in
\centerline{Figure}
\vskip .5in}
\fi

\Title{hep-th/9509132, RU-95-53}
{\centerline{\vbox{~~~Another Length Scale in String Theory?}}}
\bigskip
\centerline{Stephen H. Shenker}
\vglue .5cm
\centerline{Department of Physics and Astronomy}
\centerline{Rutgers University}
\centerline{Piscataway, NJ 08855-0849, USA}
\centerline{{\tt shenker@physics.rutgers.edu}}

\vskip 3em

\noindent
We suggest that some of the remarkable results on stringy dynamics which
have been found recently indicate the existence of another dynamical
length scale in string theory that, at weak coupling, is much shorter
than the string scale.  This additional scale corresponds to a mass
$\sim m_{\rm s}/g_{\rm s}$ where $m_{\rm s}$ is the square root of the
string tension and $g_{\rm s}$ is the string coupling constant.  In four
dimensions this coincides with the Planck mass.

\Date{9/95}
\nref\dabharv{ A. Dabholkar and J. A. Harvey,
Phys. Rev. Lett.
{\bf 63} (1989) 478.}

\nref\hetsol{ A. Strominger,
Nucl. Phys. {\bf B343} (1990) 167; Erratum: Nucl. Phys. {\bf B353} (1991)
565.}

\nref\FIQL{
A. Font, L. Ibanez, D. Lust and F. Quevedo, Phys. Lett. {\bf B249}
(1990) 35.}

\nref\DuLu{M. Duff, and J. Lu, Nucl. Phys. {\bf B354} (1991) 565;
Nucl. Phys. {\bf B357} (1991) 534.}

\nref\Sena{
A. Sen, Nucl. Phys. {\bf B404} (1993) 109, hep-th/9207053; Phys. Lett.
{\bf B303} (1993) 22, hep-th/9209016; Mod. Phys. Lett. {\bf A8} (1993)
2023,hep-th/9303057.}

\nref\Schw{
J. Schwarz, hep-th/9209125, hep-th/9307121,
hep-th/9411178, hep-th/9503127.}

\nref\ScSea{
J. Schwarz and A. Sen, Nucl. Phys. {\bf B411} (1994) 35
hep-th/9304154; Phys. Lett. {\bf B312} (1993) 105 hep-th/9305185.}

\nref\Review{
A. Sen, Int. J. Mod. Phys. {\bf A9} (1994) 3707 hep-th/9402002;
Nucl. Phys. {\bf B434} (1995) 179 hep-th/9408083;
hep-th/9503057.}

\nref\HuToa{
C. Hull and P. Townsend, Nucl. Phys. {\bf B438} (1995) 109,
hep-th/9410167.}

\nref\Duff{
M. Duff, Nucl. Phys. {\bf B442} (1995) 47, hep-th/9501030.}

\nref\Town{
P. Townsend, Phys. Lett. {\bf B350} (1995) 184, hep-th/9501068.}

\nref\wita{E. Witten, Nucl. Phys. {\bf B443} (1995) 85, hep-th/9503124.}

\nref\Senb{
A. Sen, hep-th/9504027.}

\nref\HaSt{
J. Harvey and A. Strominger, hep-th/9504047.}

\nref\str{A. Strominger, hep-th/9504090.}

\nref\gms{B. Greene, D. Morrison and A. Strominger, hep-th/9504145.}

\nref\Va{C. Vafa, hep-th/9505023.}

\nref\VaWia{
C. Vafa and E. Witten, hep-th/9505053.}

\nref\HuTob{
C. Hull and P. Townsend, hep-th/9505073.}

\nref\KaVa{
S. Kachru and C. Vafa,  hep-th/9505105.}

\nref\FHSV{
S. Ferrara, J. Harvey, A. Strominger and C. Vafa
hep-th/9505162.}

\nref\KLT{
V. Kaplunovsky, J. Louis and S. Theisen, hep-th/9506110.}

\nref\KLM{
A. Klemm, W. Lerche and P. Mayr, hep-th/9506112.}

\nref\gv{D. Ghoshal and C. Vafa, hep-th/9506122.}

\nref\PaTo{
G. Papadopoulos and P. Townsend,  hep-th/9506150.}

\nref\Dabh{
A. Dabholkar,   hep-th/9506160.}

\nref\Hull{
C. Hull,   hep-th/9506194.}

\nref\aspin{P. Aspinwall, hep-th/9507012.}

\nref\ScSeb{
J. Schwarz and A. Sen,   hep-th/9507027.}

\nref\AGNT{
I. Antoniadis, E. Gava, K. Narain and T. Taylor,
hep-th/9507115.}

\nref\VaWib{
C. Vafa and E. Witten,   hep-th/9507050.}

\nref\witb{
E. Witten,   hep-th/9507121.}

\nref\HaLoSt{
J. Harvey, D. Lowe and A. Strominger,   hep-th/9507168.}

\nref\bbs{K. Becker, M. Becker and A. Strominger, hep-th/9507158.}

\nref\senv{A. Sen and C. Vafa, hep-th/9508064.}

\nref\kklmv{S. Kachru, A. Klemm, W. Lerche, P. Mayr and C. Vafa,
hep-th/9508155.}

\nref\ap{I. Antoniadis and H. Partouche, hep-th/9509009.}

\nref\suss{L. Susskind, hep-th/9409089, and private communications.}

\nref\grossmende{D. J. Gross and P. Mende, Nucl. Phys. {\bf B303} (1988)
407.}

\nref\seibrev{For reviews see N. Seiberg in PASCOS 1994 hep-th/9408013,
and in PASCOS 1995 hep-th/9506077.}

\nref\sw{N. Seiberg and E. Witten,
Nucl. Phys. {\bf B426}, (1994), 19, hep-th/9407087}

\nref\ds{M. R. Douglas and S. H. Shenker, Nucl. Phys. {\bf B447},
(1995), 271, hep-th/95031633.}

\nref\afcern{P. Argyres, A. Faraggi, Phys. Rev. Lett. {\bf 74} (1995)
3931,
hep-th/9411057; A. Klemm, W. Lerche, S. Yankierlowicz and S. Theisen,
Phys. Lett. {\bf B344} (1995) 169, hep-th/9411048.}

\nref\deWit{B. de Wit, P. Lauwers and A Van Proeyen,
 Nucl. Phys. {\bf B255} (1985) 269.}

\nref\mir{P. Candelas, X. de la Ossa. P. Green and L. Parkes,
Nucl. Phys. {\bf B359} (1991) 21. }

\nref\shs{S. H. Shenker, in {\it Random Surfaces and Quantum Gravity},
O. Alvarez, E. Marinari, and P. Windey (eds.), Plenum (1991) 191.}

\nref\bcov{M. Bershadsky, S. Cecotti, H. Ooguri and C. Vafa, Comm. Math. Phys.
{\bf 165} (1994) 311, hep-th/9309140; Nucl. Phys. B{\bf405} (1993) 279.
}

\nref\mv{S. Mukhi and C. Vafa, Nucl. Phys. {\bf B407} (1993) 667,
hep-th/9301083.}

\nref\gm{G. Moore, Nucl. Phys. {\bf B368} (1992) 557. For reviews of
matrix models see e.g.,  I. Klebanov in PASCOS 1991, World Scientific,
(1992) 526; P. Ginsparg and G. Moore in TASI 1992, hep-th/9304011.}

\nref\hs{G. Horowitz, and A. Strominger, Nucl. Phys. {\bf B360} (1991) 197.}

\newsec{Introduction}
Our understanding of the dynamics of string theory has dramatically
increased recently \refs{\dabharv-\ap}.  We now understand, via strong
coupling-weak coupling dualities, that apparently different theories are
different regions of the same theory.  We  also understand that
apparently disjoint vacuum states are in fact connected.

The existence of
certain kinds of solitons is crucial to these dynamical insights.
At weak string coupling, $g_{\rm s}^2$, these solitons generically
have masses $m_{\rm sol}\sim m_{\rm s}/g_{\rm s}^2$, or for those carrying
Ramond-Ramond (RR) charge, $m_{\rm sol}\sim m_{\rm s}/g_{\rm s}$ \wita .  Here
$m_{\rm s}$ is the basic string mass scale $m_{\rm s} \sim \sqrt{T}$
where $T$ is the string tension.  When the coupling is weak
these solitons are much heavier than $m_{\rm s}$.

This by itself is not surprising.  In field theory, even though solitons
define a new heavy mass scale, this scale does not have dynamic
consequences at short distance or high momentum.  The basic reason for
this is that solitons, while heavy, are also big.  As an example
consider a nonabelian gauge theory spontaneously broken to $U(1)$.
There are magnetic monopoles in this theory whose mass at weak coupling,
$g^2$, is $m_{\rm mon}\sim m_{\rm W}/g^2$.  Here $m_{\rm W}$ is the
mass of a typical massive gauge vector boson.  Even though $m_{\rm mon}
\gg m_{\rm W} $ we do not expect the short distance behavior of this
gauge theory to be anything but that of a theory of weakly interacting
gauge bosons.  $1/m_{\rm W}$ is the shortest dynamical length scale in the
problem.  It is also the semiclassical ``size'' of the monopole, far
larger than $1/m_{\rm mon}$.

In string theory we have come to believe that the shortest
dynamical length scale in the theory is $1/m_{\rm s}$. There are many
indications of this, including gaussian fall-off of
high momentum fixed angle scattering, Regge behavior,
$R \rightarrow 1/R$ duality,
and the Hagedorn transition.  Because of the field theory intuition
just mentioned the existence of heavy solitons has not caused us to
question this conventional wisdom.

The aim of this paper is to suggest
that, on the contrary, some of the new information on string dynamics
seems to indicate that $m_{\rm s}/g_{\rm s}$ serves as another  {\it dynamical}
scale in string theory.  The evidence we will present for this
additional scale is quite indirect.  It may be that there is an
explanation for it that does not require the existence of another
scale.  Nonetheless the evidence is sufficiently puzzling that it
seems worthwhile to present it here.

This is not the first suggestion of another dynamical scale in string
theory.  Susskind has argued \suss\ that at energies above the Planck mass
$m_P$ ($\sim m_{\rm s}/g_{\rm s}$ in four dimensions) strings must become black
holes and behave rather differently than perturbative strings.  He
points to the flattening of the gaussian falloff of fixed angle
scattering as the order of perturbation theory is increased \grossmende\
as an indication of a possible change in high momentum behavior.

\newsec{Logarithms}

\seclab\logs

The evidence for another length scale comes from new
exact results about certain long distance quantities in low
energy effective Lagrangians.
Long distance quantities can give some information about short distance
behavior because of the high momenta present in loops of virtual
particles that contribute to them.

To see how this works let us examine a field theoretic example.  A large
number of exact low energy results in supersymmetric field theory are
now available due to the enormous recent progress in this subject
spearheaded by Seiberg \seibrev.

To be explicit we will consider the exact low energy effective action of
$\CN=2$ $SU(2)$ supersymmetric gauge theory determined in the beautiful
work of Seiberg and Witten ~\sw.  The low energy effective dual U(1)
coupling constant $\tau_D = 4\pi i/e_D^2 + \theta/2\pi$ depends on the
modulus of the theory, the complex valued gauge
invariant vacuum expectation value of the adjoint scalar field.  Let
$Z=0$ be the point of this moduli space where monopoles become massless.
Near this point

\eqn\tauSW{\tau_D(Z) \sim {i\over{2 \pi}} \log (Z) ~.}

This singular behavior is explained in \sw\ as the infrared divergent charge
screening due to a light
magnetic monopole.  Such an explanation requires interpreting
the logarithm in \tauSW\
as $ -\log (m_{\rm uv}/m_{\rm mon}) $ where $m_{\rm mon}$ is the mass of the
magnetic monopole in the theory and $m_{\rm uv}$ is the
ultraviolet cutoff in the loop integral.  We might might think of
$m_{\rm uv}$ as the inverse ``size'' of the monopole. It represents
the scale at which a description purely in terms of light pointlike
monopoles breaks down and the logarithmic functional form in \tauSW\
ceases to be accurate.
The size of $m_{\rm uv}$ is a piece of
knowledge about short distance physics gained from long distance measurements.

In theories with extended supersymmetry, states in reduced multiplets
obey a BPS mass formula.  For the theory in \sw\ as $Z \rightarrow 0$
this formula give $ m_{\rm mon} \sim |Z|\Lambda$ and $m_{\rm W} \sim
\Lambda$ where $\Lambda$ is the asymptotic freedom mass scale. Using
these relations we see that $m_{\rm uv} \sim m_{\rm W}$ up to factors of
order one, in accord with semiclassical intuition about the ``size'' of
a monopole.  This interpretation is bolstered by the analysis of the
$SU(N)$ generalization of the Seiberg-Witten solution \afcern\ given in
\ds.  In the $SU(N)$ case there are $N-1$~ $U(1)$ gauge fields at low
energy. At large $N$ there is a large hierarchy of different length
scales in the theory at the massless monopole point.  The $W$ masses
there range from $\sim \Lambda$ to ${1 \over N^2} \Lambda$.  For each
$U(1)$ factor (indexed by $a$) the analog of  \tauSW\ ceases to be
a good description when $m_{\rm mon_a}$ becomes $\sim m_{\rm uv_a} \sim
m_{\rm W_a}$ where $m_{\rm W_a}$ is the mass of the lightest charged
vector boson that couples to $U(1)_a$.
This all seems very
sensible.

Let us now turn to string theory and in particular examine Strominger's
remarkable resolution of the conifold singularity~\str.  He studies 4D
type IIB superstring theory compactified on a Calabi-Yau manifold (CY) that
develops a conifold singularity as a modulus denoted by $Z$ goes to
zero.  This theory has $\CN = 2$ 4D supersymmetry and $Z$ is the scalar
component of a vector multiplet.  The $U(1)$ vector gauge field in this
multiplet is a RR field.  Its effective low energy coupling constant
(related by supersymmetry to the geometry of moduli space) has been
computed exactly at string tree level using mirror symmetry \mir.  This
result should be exact in quantum string theory \str\ because the IIB
dilaton lies in a neutral hypermultiplet which cannot couple to the
vector multiplets by $\CN=2$ supersymmetry.  The result for the
effective coupling $\tau_{RR}$ as $Z \rightarrow 0$ is precisely \tauSW:

\eqn\tauRR{\tau_{RR}(Z) \sim {i\over{2 \pi}} \log (Z)~.}

Strominger explains this singularity\foot{ In the limit $m_{\rm bh}
\ll m_{\rm s}$.}
as the screening due to an (electrically) charged light gravitational
soliton referred to as a ``black hole.''  Again this explanation
requires interpreting the logarithm in
\tauRR\ as $ -\log (m_{\rm uv}/m_{\rm bh}) $ where $m_{\rm bh}$
is the mass of the black hole and $m_{\rm uv}$ is the ultraviolet cutoff
in the loop integral.  The BPS formula for this RR charged state gives,
for $Z
\rightarrow 0$,

\eqn\bpsbh{m_{\rm bh} \sim |Z| m_{\rm s}/g_{\rm s} ~.}

This implies that here $m_{\rm uv} \sim m_{\rm s}/g_{\rm s}$ !  If
$m_{\rm uv}$ were $\sim m_{\rm s}$ then $\tau_{RR}$ would be
$\sim \log(Z/g_{\rm s})$
implying a coupling between RR vector multiplets and the neutral dilaton
hypermultiplet $D$,  ($g_{\rm s} \sim e^{D}$), and in particular a nonvanishing
three point vertex between two RR vector bosons and a zero momentum
dilaton.  Such couplings are forbidden by $\CN=2$ supersymmetry \deWit\
\str .

Somehow this black hole seems to be behaving like a pointlike four
dimensional particle down to length scales many times smaller
(for small $g_{\rm s}$) than either string or compactification lengths.
This does not seem very sensible.

Even before we address string scales there is a problem.  At some
compactification mass scale $m_{\rm c}$ (which could be much less than
$m_{\rm s}$) the theory stops looking like a four dimensional theory and
starts looking ten dimensional.  Why do we keep doing a four dimensional
loop momentum integral?  One possible explanation for this \foot{This
explanation was suggested to me by Tom Banks and Andy Strominger.  My
understanding of it was shaped by conversations with them and with Jeff
Harvey, Emil Martinec and Greg Moore. } is that the black hole soliton
is ``locked'' in position on the six dimensional CY.

Perturbative
orbifold compactified string theory displays an analogous locking
phenomenon.
Consider an orbifold compactification from ten to four dimensions where
the the radius of the orbifold is $R = 1/m_{\rm c}$ and assume $m_{\rm c} \ll
m_{\rm
s}$.  In the untwisted sector there are massless states and Kaluza-Klein (KK)
excitations above them of mass $\sim m_{\rm c}$ (as well as stringy
excitations). These KK states cause the theory to look ten dimensional
at energies above $m_{\rm c}$.  In the twisted sector, though, there are no KK
excitations.  In world sheet language, there are no zero modes of the $X$
fields and the only excitations come from oscillators.  In spacetime
language, any attempt to move the string away from the orbifold
singularity on which it is trapped changes its length and hence costs string
tension scale energy.

Strominger \str\ argues that the black hole should be thought of as a
threebrane wrapped around the collapsing three-cycle that defines the
conifold singularity.  It is plausible that any
excitation of this wrapped threebrane away from the three-cycle
will change the world volume and hence cost brane tension energy and be very
massive. This would be locking.

Perhaps there is an indirect (and partially circular!) argument for
locking along the following lines.  If the black hole were not locked
its ability to move in the six compact dimensions would be
represented by charged KK modes in the 4D field theory whose mass was $\sim
m_{\rm c}$. But $m_{\rm c}$ depends on Kahler moduli which are part of neutral
$\CN=2$
hypermultiplets.  Integrating out such fields\foot{Which should,
as Emil Martinec pointed out, not
in general be part of reduced $\CN=2$ multiplets.  Andy Strominger
noted that multiplets consisting of paired vector and hypermultiplets give
zero contribution to $\tau_{RR}$.  Perhaps we need to consider
higher spin multiplets.}
should give Kahler moduli
dependence to $\tau$ which is forbidden by $\CN=2$.

Locking might explain why $m_{\rm uv}$ is much larger than $ m{\rm c}$.
The fact that $m_{\rm uv}$ is much larger than $ m_{\rm s}$ seems a much
deeper mystery.  How can the theory still display pointlike particle
behavior at distances much shorter than the string scale?  Why doesn't
the theory dissolve into soft mushy strings?  We will face this question
repeatedly in this paper.  One hint comes from an analogy to the
Seiberg-Witten case.  There $m_{\rm uv}$ was determined by the mass of a
field carrying $U(1)$ charge.  There are no perturbative string states
carrying RR charge, only solitons.\foot{Tom Banks first stressed to me
that this may be related to the lack of a string scale cutoff.} The
lightest such state, aside from the light black hole, has mass $\sim
m_{\rm uv}$.

\newsec{Four to Three Dimensions}

\seclab\ftt

At first glance there seems to be a serious obstacle to interpreting the
conifold logarithm as charge renormalization due to a light black hole.
This logarithm is computed in tree level conformal field theory, whose
target manifold factorizes into $\CM \times {\IR}^4$ where $\CM$ is the
CY manifold.  The logarithm comes entirely from the $\CM$ factor which
is independent of $\IR^4$. In particular if we replace $\IR^4$ with
$\IR^3 \times S^1$ where the radius of the $S^1$ is $R$ we find an
unmodified logarithm.  But if the logarithm is reflecting a four
dimensional infrared divergence how can it not be sensitive to
compactifying one of the four dimensions?  At length scales large compared to
$R$ the infrared behavior should be three dimensional, not four!  But
$R$ does not appear in the effective coupling.

This puzzle also seems soluble if we assume that $m_{\rm uv}
\sim m_{\rm s}/g_{\rm s} $.\foot{This observation was motivated
by Ed Witten's remark that in Einstein frame the radius of the
compact dimension is $R/g_s$ and hence goes to infinity as $g_s$ goes to
zero.  So the effects of compactification should disappear at weak
coupling.}  To see how
this might work, consider the following model loop integral\foot{This
calculation was developed in a discussion with Ronen Plesser.} for the
effective coupling in four dimensions, $e^2_{4D}$.

\eqn\tfour{1/e^2_{4D} \sim \int d^4 p {1 \over{(p^2 + m_{\rm bh}^2)^2}}
{{m_{\rm uv}^2} \over {(p^2+m_{\rm uv}^2)}} \sim
\log(m_{\rm uv}/m_{\rm bh}) ~.}

The second factor in the integrand ensures a smooth high momentum
cutoff at $m_{\rm uv}$.  Its detailed form is unimportant. In $\IR^3 \times
S^1$ the model becomes

\eqn\tthree{1/e^2_{3D} \sim {1 \over R }\sum_{n} \int d^3 p
{1 \over{(p^2 + ({{2 \pi n} \over
{R}})^2 + m_{\rm bh}^2)^2}} {{m_{\rm uv}^2} \over {(p^2+({{2 \pi n} \over
{R}})^2 +m_{\rm uv}^2)}} ~.}

If $1/R \ll m_{\rm bh}$ then we have, schematically,
\eqn\tR{1/e^2_{3D} \sim 1/e^2_{4D} + \e{{- m_{\rm bh} R}} +\e{{-2 m_{\rm bh}
R}} + \ldots
+\e{{-m_{\rm uv} R}} +\e{{-2 m_{\rm uv} R}} + \ldots ~.}

This follows from interpreting large $R$ as low temperature or, more
formally, by Poisson resummation of \tthree.  We now look at the finite
$R$ corrections in \tR.  Using ~\bpsbh\ we see that the term
$\exp(-m_{\rm bh} R)$ becomes $\exp(-{{|Z|m_{\rm s} R}\over{g_{\rm
s}}})$.  For $Z$ and $m_{\rm s} R$ fixed this term is nonperturbative in
$g_{\rm s}$.  In fact it is an example of the mechanism suggested
in \wita\ for producing the nonperturbative effects of ``stringy''
strength argued to exist in general in \shs\ on the basis of the large
order behavior of string perturbation theory.  It is also an example
of the instanton effect discussed in \bbs.  The instanton here is the
black hole circling around the $S^1$.

Because this $R$ dependence is nonperturbative in $g_{\rm s}$ it does not
contradict the factorization noted above at tree level.  The
distinction between vector and hypermultiplets disappears in three
dimensions so we expect nonperturbative corrections to $\tau_{RR}$ \bbs.

Now consider the term $\exp(-m_{\rm uv} R)$.  For $m_{\rm uv} \sim
m_{\rm s}$ this term has tree level coupling dependence.  This
apparently does contradict
factorization.  But if $m_{\rm uv} \sim m_{\rm s}/g_{\rm s}$ then this
term is $\sim\exp(-{{m_{\rm s} R}\over{g_{\rm s}}})$.  This
has nonperturbative $g_{\rm s}$ dependence so again
there is no problem.  In fact the term is then of the same form as
that due to other instantons discussed in \bbs\ corresponding to heavy
solitons circling the $S^1$, making more plausible the identification of
$m_{\rm uv}$ with a heavy soliton mass.

Compatibility with factorization does not really require that $m_{\rm
uv}
\sim m_{\rm s}/g_{\rm s}$.
A weaker $g_{\rm s}$ dependence, e.g., $m_{\rm uv} \sim
m_{\rm s}/\sqrt{g_{\rm s}}$ would suffice. This would correspond to stronger
nonperturbative effects $\sim\exp(-{{1}\over{\sqrt{g_{\rm s}}}})$
and hence a large order behavior faster
than $(2g)!$ at genus $g$, which would be surprising.

\global\advance\ftno by1

\newsec{Dual Picture$^{\the\ftno}$}
\vfootnote{$^{\the\ftno}$}{The observations in this section were motivated
by Ashoke Sen's and Cumrun Vafa's remark that the scale $m_{\rm
s}/g_{\rm s}$ is the natural one in the dual heterotic description
of the conifold. }

\seclab\dualp

In the last year or so, strong evidence has accumulated for the validity
of various exact string-string dualities \refs{\dabharv-\ap}.  We can
ask whether these dualities shed some light on the length scale question
we have raised.  Kachru and Vafa \KaVa\ proposed  duals for  4D $\CN=2$
type II strings compactified on certain CY manifolds having loci of
conifold singularities in their moduli spaces.


One of the duals they found was a 4D $\CN=2$ heterotic string
compactified in a such a way that at tree level there is a point of
enhanced $SU(2)$ gauge symmetry in its moduli space.  The flat space low
energy effective field theory near this point is just $SU(2)$ $\CN=2$
gauge theory, the model analyzed by Seiberg and Witten\sw.  Seiberg and
Witten have taught us that quantum mechanically the $SU(2)$ point splits
into two singular points where magnetic monopoles or dyons become
massless.  The dual image of these points on the type II side are
conifold singularities!
\foot{There are two such loci in this example, as required.}
The type II dual image of a
heterotic magnetic monopole becoming massless is an electrically charged
black hole becoming massless at a conifold point.  The accompanying
figure gives a description of the spectrum in both the heterotic and
type II languages in the $Z \ll g^{\rm II}_{\rm s}
\ll 1$ regime.

\epsfxsize=\hsize
\fig{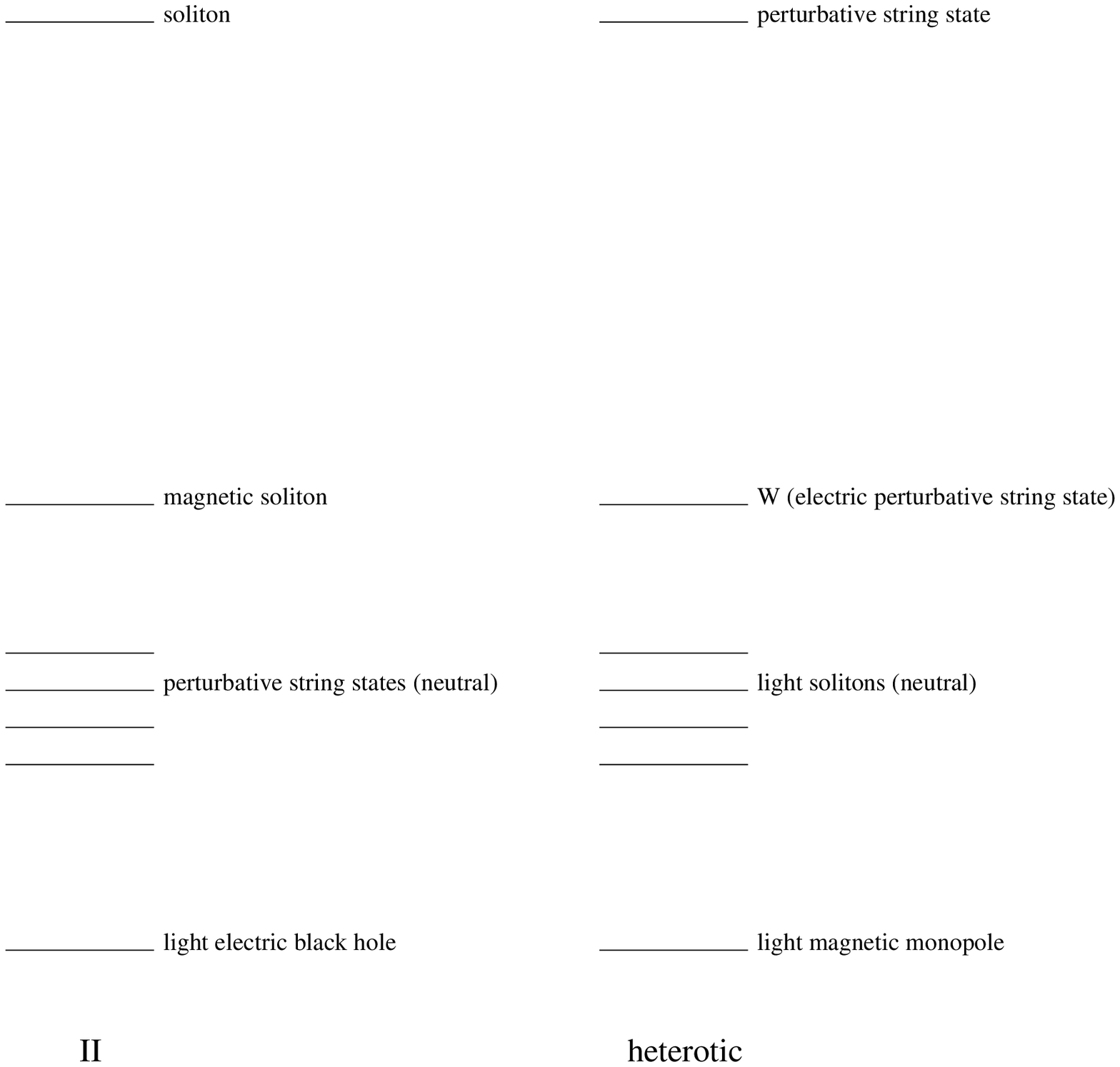}{}

By using the magic of second quantized mirror
symmetry \FHSV \KaVa\ the stringy analog of
\tauSW\ on the heterotic side
can be computed from nonperturbatively exact tree level results on the
type II side.  Assuming that the physics on the heterotic side is
essentially that of \sw\ we can determine $m_{\rm uv}$.
As discussed in section \logs\ it should just be
$m_{\rm W}$ at the massless monopole point.  For flat space physics to
be accurate $m_{\rm W}$ should be much less than the heterotic string
scale $m_{\rm s}^{\rm het}$, but only by a fixed ratio, independent,
e.g., of the value of $g_{\rm s}^{\rm II}$ on the type II side \kklmv.  But
$m_{\rm W}$ carries RR charge, so its type II image is a heavy
nonperturbative (magnetically) charged soliton of mass $\sim m_{\rm
s}^{\rm II}/g_{\rm s}^{\rm II}$. This corresponds
to the guess about the identity of $m_{\rm uv}$ made in the
previous two sections.  The mystery from the heterotic point of view
concerns the heterotic images of the neutral perturbative type II
states. These states have mass $\sim m_{\rm s}^{\rm II} \sim g_{\rm
s}^{\rm II} m_{\rm s}^{\rm het}$ and so are much lighter than the
perturbative heterotic states.\foot{ The heterotic theory is strongly
coupled so this is not implausible, and of course is required by
duality.} Why don't they serve to cut off the logarithm?  The exact answer
shows no dependence on their mass, since the answer only depends on heterotic
vector multiplets.  Again,
part of the explanation may be that they are neutral.

\global\advance\ftno by1

\newsec{Finite Momentum$^{\the\ftno}$}
\vfootnote{$^{\the\ftno}$}{Much of this section was developed in various
discussions with Misha Bershadsky, Mike Douglas, Daniel Friedan, Hirosi
Ooguri, Andy Strominger and Cumrun Vafa.}

\seclab\finm

So far the signatures we have discussed for physics at scales
$\sim m_{\rm s}/g_{\rm s}$ are quite indirect.
They occur in low energy quantities and all depend in one way
or another on the $\CN=2$ supersymmetric decoupling of vector
and neutral hypermultiplets.
If such a scale is important
in string dynamics we would expect to see direct evidence for it
in high momentum scattering.  To take an extreme
example, charged black hole--charged black
hole scattering should show some feature at momenta $\sim m_{\rm s}/g_{\rm s}$.

We are a long way from being able to calculate a quantity like this.
But we can find some hints about how other momentum scales besides
$m_{\rm s}$ can enter into results.  Consider the scattering of two RR
$U(1)$ vector bosons near a conifold degeneration.  The effective low
energy physics should contain a light black hole field, and so we should
see a threshold for black hole pair creation. There should be a feature
like $(p^2+m_{\rm bh}^2)^{\alpha}$ in the S-matrix.  Expanding such a term in
powers of $p^2$ gives a series that looks schematically like
\eqn\seriesm{
S(p)\sim \sum_l (p^2/m_{\rm bh}^2)^l ~.}
Using \bpsbh\ we can rewrite this as
\eqn\seriesg{S(p)\sim \sum_l (p^2 g_{\rm s}^2/Z^2)^l ~.}
So a possible signature of this threshold would be a $(p^2/Z^2)^l$
behavior at genus $l+1$ in string perturbation theory as $Z \rightarrow 0$.
We also expect terms like
$(p^2/m_{\rm bh}^2)^l (m_{\rm s}^2/m_{\rm bh}^2)^n
\sim p^{2l}(g_{\rm s}^2/Z^2)^{n+l}$.  These will be quantitatively important
since the simple black hole picture will be valid when $m_{\rm bh} <
m_{\rm s}$.  The important point is that no inverse powers of $Z$
appear without accompanying $g_{\rm s}$'s.

We can identify a potential source of such terms.  Consider the four Z
moduli scattering amplitude $A_{ZZZZ}$ (related by supersymmetry to the
four vector amplitude) at tree level.  This amplitude is related to the
curvature of the metric on moduli space defined by the effective $U(1)$
gauge couplings, including $\tau_{RR}$. Differentiating \tauRR\ twice we
find
\eqn\AZZZZ{A_{ZZZZ}(p) \sim p^2 g_{\rm s}^2/Z^2 ~.}
Repeated iterations of this tree level four moduli  scattering  could
produce terms of the type found in \seriesg.
Because the loop momenta in such a diagram
are integrated over, a variety of different
terms with different powers of momenta
as mentioned above
could result.

The above arguments are very similar to scaling arguments presented in
\bcov .  In \bcov\ the authors study the coefficients $C_l$ of certain
terms in the effective action of the form $F^{2l}R^2$ where $F$ is the
RR field strength and $R$ is the Riemann tensor.  They showed that such
a term only gets a contribution from genus $l+1$ in string perturbation
theory and that in the conifold limit $Z \rightarrow 0, ~C_l \sim (g_{\rm
s}^2/Z^2)^l$ with scaling like the terms in \seriesg.  The higher powers
of momentum in \seriesg\ are replaced by nonrenormalizable operators with
many derivatives and many RR gauge fields.

Ghoshal and Vafa \gv\ made further progress by connecting the conifold
with a Kazama-Suzuki $SL(2)/U(1)$ coset.  This latter,
after topological twisting, had already been shown to be equivalent
to the $c=1$ noncritical string  by Mukhi and Vafa \mv.
The result of this series of mappings is the fascinating result
that the $C_l$ are precisely the coefficients of the $c=1$ matrix model
partition function at the self dual radius! \foot{These $C_l$ have
been computed in the dual heterotic picture in the beautiful paper
\AGNT. As mentioned in \AGNT\ terms like $F^{2l}$ would naturally
occur in studying the perturbation expansion of the theory in a uniform
RR field strength.  Perhaps the nonperturbative effects associated with
the $c=1$ model are, in this context, to be interpreted as pair
production of light black holes in a background field.  The
identification of $\mu$ with the black hole mass discussed later in this
section supports this.  This idea was developed in a discussion with
Mike Douglas and Andy Strominger and is also implicit in \AGNT.}

The work of Witten \witb\
motivates a possible physical picture for this occurrence.
Witten
has argued that in the conifold limit (and in other related
degenerations) the quantum corrected geometry of the target manifold
develops a long tube whose length goes to infinity as $Z \rightarrow 0$.
The dilaton varies linearly along the tube.  Such linear dilaton
backgrounds are characteristic of noncritical strings. Conjecturally
the tube, after topological twisting, will be the $c=1$ background.\foot
{This observation was developed in discussions with Hirosi Ooguri and
Cumrun Vafa.  We hope to report on further developments in the not
too distant future.}
The ``double scaling limit'' of the theory, $g_{\rm s} \rightarrow 0, ~
Z \rightarrow 0, ~\mu \equiv g_{\rm s}/Z$ held fixed, extracts the physics of
the tube.
Here $\mu$ is to be identified with the
continuum coupling (also denoted by $\mu$))  of the $c=1$ matrix model.
It is very interesting to observe that $\mu$ is just the mass of the
black hole in string units, $\mu = m_{\rm bh}/m_{\rm s}$.  The double scaling
limit of the conifold holds the mass of the black hole fixed while
the string coupling is taken to zero.  It may well be that this limit
provides a simpler model in which to study certain aspects of the physics
of these light black holes.

Here we content ourselves with the following vague observations about
higher energy scales.  The
$c=1$ model has nonstandard large momentum behavior.  At genus g
``tachyon'' scattering amplitudes grows like $(k^2/\mu)^{2g}$ at large
momentum $k$\gm.  Here $k$ is the momentum along the direction of linear
dilaton growth measured in string units.  $\mu$ again serves as a
characteristic scale indicating when new physics sets in and
perturbation theory breaks down.  In this model the new physics is the
possibility of fermions going over the potential barrier of height
$\mu$.  There is a simple picture for this kind of momentum dependence.
The higher the momentum down the tube, the deeper the particle can
penetrate into the tachyon condensate ``wall'' and hence experience a
stronger coupling.  Perhaps a related high momentum behavior occurs in
conifold amplitudes.  The new physics signaled by this behavior would
presumably
have to do with black hole creation, as black holes have mass $\mu $ in
string units.
Such behavior would signal the breakdown of string perturbation theory
and perhaps its associated soft high momentum behavior.

\newsec{Some Puzzles}

There are many puzzles posed by the possible existence of this new scale.
In this section we will discuss some of them.

We have argued that interpreting \tauRR\ as the result of a  one loop
vacuum polarization graph, where a low momentum RR gauge field excites a
virtual charged black hole pair, requires that the loop momentum of the
black hole run up to $\sim m_{\rm s}/g_{\rm s}$.
But the RR gauge field is a conventional
perturbative string state which we might have thought would have
``size'' $1/m_{\rm s}$.  How can it couple in a pointlike way to the charged
black hole all the way up to momenta of order $\sim m_{\rm s}/g_{\rm s}$?
\foot{
This puzzle arose in a discussion with Lenny Susskind.} We might be able
to think of the structure of the RR vector state
as coming from the exchange of additional neutral
perturbative string states in the diagram.  We don't really know
how to work with such diagrams, since we don't know the coupling
of perturbative string states to light black holes, and the coupling
of the black holes to RR vectors is order one.  Nonetheless we might
be able to attribute the absence of string scale effects in $\tau_{RR}$
to a cancellation analogous to that responsible for the $\CN=2$
nonrenormalization theorems that prevent neutral hypermultiplets from
contributing here.
Such exchanges would affect other
quantities and suggest a string scale ``halo'' surrounding these
solitons that special quantities like $\tau_{RR}$ don't see.  The
ultraviolet cutoff at $m_{\rm s}/g_{\rm s}$ would correspond to a much
smaller ``charge radius'' of the soliton.

The
$\exp(-1/g_{\rm s})$ nonperturbative effects expected at weak coupling \shs\
are another indication of such a ``halo.''
World line
loops of RR solitons were proposed as a mechanism for such effects in
\wita.  In order for the action of such a loop to be $\sim 1/g_{\rm s}$
the loop's size must be $\sim 1/m_{\rm s}$.
If there is no suppression
of smaller loops then the contribution of these solitons will only be
power law suppressed by inverse powers of their mass.\foot{
These points were made by Lenny Susskind.}
This suggests a
``size'' for the solitons $\sim 1/m_{\rm s}$.
Supersymmetry nonrenormalization theorems
do not prohibit perturbative string state exchanges from affecting,
e.g., the coefficient  $c_4$ of $F^4$
where $F$ is the RR field strength.  The general arguments of
\shs\ would suggest $(2g)!$ large order growth in the perturbation
series of $c_4$ since it is not protected by supersymmetry.
We would then expect $\exp(-1/g_{\rm s})$ corrections to $c_4$.
Since the string state exchanges do not cancel they could stabilize
the minimal loop of RR solitons coupling to such a quantity at a length
$\sim 1/m_{\rm s}$, giving the requisite action.  An important test
of such a picture would be to isolate the nonperturbative effects contributing
to such a quantity from the light solitons of mass $|Z|m_{\rm s}/
g_{\rm s}$.  These would
be of order $\exp(-Z/g_{\rm s})$ assuming a minimum loop size $\sim
1/m_{\rm s}$.  The corresponding large order perturbative behavior
of a four vector scattering amplitude at low momentum $p$ and $Z
\rightarrow 0$
would look schematically like
\eqn\seriespuz{S(p)\sim p^4\sum_l (g_{\rm s}^2/Z^2)^l (2l !)~.}
Note the contrast with \seriesg.  Understanding
the full momentum dependence of the double scaling
limit of such scattering amplitudes will be instructive.

If, at weak coupling, the solitons should be thought of as having a
a ``halo'' $\sim 1/m_{\rm s}$ perhaps all direct effects of their
existence will be exponentially suppressed.

In the type II language the conifold phenomenon is in some sense a weak
coupling one, although some perturbative states are certainly strongly
coupled when $m_{\rm bh} \rightarrow 0$ even though
$g_{\rm s}$ can be arbitrarily small.  Nonetheless we might expect to get
some insight into the occurrence of the $m_{\rm s}/g_{\rm s}$ scale by
examining the classical solution for the black hole.\foot{I thank Andy
Strominger for teaching me most of what follows.}

Strominger has argued that the appropriate classical configuration is
the threebrane of the 10D IIB theory \hs wrapped around the shrinking
three-cycle of the CY. The threebrane has a ``throat'' whose width defines
a characteristic size.  We might expect this size to be string scale,
but in fact it is not.  The threebrane only carries RR charge
(described by the integral of the self-dual five-form) and so the
peculiarities of RR charge quantization come into play.  The dilaton
can be completely eliminated from the Lagrangian by going to Einstein
frame and rescaling the
RR fields so they are quantized in unit strength.  The size must be,
then, the only scale remaining, which is the Planck scale.
Note that this is the ten dimensional
Planck scale $m_P^{\rm (10)} \sim m_{\rm s}/g_{\rm s}^{1/4}$, {\it not} the
four
dimensional Planck scale $m_P^{\rm (4)} \sim m_{\rm s}/g_{\rm s}$.
Other solitons that do carry NS charge, for example the symmetric
fivebrane
used to construct the heterotic soliton~\Senb\HaSt, have throat sizes
that are string scale.

What apparently is missing in this classical description is the analog
of the W bosons discussed in section \dualp\ that seem to be providing
the physical cutoff for RR charge fluctuations.

It will be important to look at this scale issue in various dimensions,
(e.g., for the massless black holes that appear at the degenerations of
K3 in 6D type II string theory), to discriminate between $m_P$ and
$m_{\rm s}/g_{\rm s}$ which happen to coincide in four dimensions.  The
mass of RR solitons is always $\sim m_{\rm s}/g_{\rm s}$, independent of
dimension \wita.

As an aside, let us point out that the peculiar quantization of RR
charge gives an intuitive explanation for the unusual lightness of RR
solitons as compared to conventional solitons whose mass is $\sim m_{\rm
s}/g_{\rm s}^2$.  The low energy field theory measure with fields
defined in units natural to string theory looks schematically like
$\exp(-S/g_{\rm s}^2)$ where S is the action.  Now typical classical
field configurations corresponding to solitons have field magnitudes of
order unity and so $S
\sim 1$.  The typical soliton mass
(derived from the action for a world line) is then $\sim m_{\rm s}/g_{\rm
s}^2$.
But the RR quantization condition in string units requires that the RR
field strength $F \sim g_{\rm s}$ and so the RR field is much weaker than
unity.  This allows the action to be \foot{In fact, the bulk
action is zero and the mass is given by a surface term.  I thank Tom
Banks for a discussion about this.} $\sim g_{\rm s}$
and the soliton mass to be $\sim m_{\rm s}/g_{\rm s}$.\foot{This is
reminiscent of the one eigenvalue instanton mechanism that produces
$\exp(-1/g_{\rm s})$ effects in matrix models.}

Another important puzzle concerns the limit $g_{\rm s} \rightarrow 0$,
$Z$ small but fixed.  In this limit string perturbation theory is
accurate and so there must be a conventional conformal field theory
explanation for the logarithm in \tauRR.  As discussed in \bcov\Va\witb\
the CY conformal field theory produces many perturbative string states
going to zero mass as $Z \rightarrow 0$ which have couplings $\sim g_{\rm
s}/Z$.  These states, in some sense, account for the logarithm.
As $Z \rightarrow 0$
with $g_{\rm s}$ fixed the black hole mass drops below $m_{\rm s}$, the
light perturbative states become strongly coupled and the description in
terms of light black holes becomes the only simple one available.  But
somehow these two descriptions must be connected.  The light
perturbative string states should be related to black holes.  These states
are neutral (as are all perturbative states) so one might guess that
they are some kind of black hole-anti black hole composites.  They would
remain light as the black holes became heavy because the black holes
interact strongly with each other.  The Sine-Gordon model might provide
an analogy. There the solitons form bound states called breathers.  As
$\hbar \rightarrow 0$ the low lying breathers remain light even though
the solitons become infinitely heavy.  Motivated by the connection
mentioned in section
\finm, we might think of the $c=1$ matrix model as another analog.  The
very light ``tachyons'' are particle-hole excitations of the fermi
surface.  There are other excitations of energy $\mu (= m_{\rm bh})$
corresponding to particles flying over the barrier.  As $\mu \rightarrow
\infty$  the ``tachyons'' remain light.  To complete the story we would
have to understand how the one loop diagram evolves into some
kind of bound state exchange as $g_{\rm s} \rightarrow 0$.

Another question raised by the existence of a short scale concerns finiteness.
We are used to thinking that the softness of string theory at momenta
higher than $m_{\rm s}$ is crucial to the theory's good ultraviolet behavior.
But if there is another much shorter dynamical length scale in the
problem how do we know that the theory still stays well behaved?

Now we
know that even close to the conifold limit, string perturbation theory
is finite order by order in $g_{\rm s}^2$.  So for example we know that the
log is cut off.  To what extent does this assure us that the full theory
is uv finite?

Finally we can ask about this short length scale far away from the conifold
limit, say $Z \sim 1$.  At weak coupling there is nothing unusual about
the spectrum--the light states discussed above have moved up to
string scale.  Do they still represent black hole bound states?
Are all direct effects of black holes at low momenta suppressed by
$\exp(-1/g_{\rm s})$ ?

Of course the biggest question raised by the existence of another
 short length scale is
what lesson it is trying to tell us about the correct formulation of
string theory.

\newsec{Concluding Remarks}
We have seen that several puzzles --
that $\tau_{RR}$  behaves as $\log(Z)$ and not $\log(Z/g_{\rm s})$,  that
$\tau_{RR}$ is essentially independent of $R$ after compactification,
and aspects of the dual mapping of the conifold singularity onto the
Seiberg-Witten massless monopole point--can be understood by positing
the existence of another dynamical scale
$m_{\rm uv} \sim m_{\rm s}/g_{\rm s}$.
On the other hand the existence of such a scale raises several puzzles
of its own.

To better understand these issues it will be important to assemble more
evidence for this scale, especially in dimensions other than four where
$m_P \neq m_{\rm s}/g_{\rm s}$.  Hopefully we can find evidence of a
more direct nature whose form is not so strongly constrained by
supersymmetry.

As we search, though, we must bear in mind the
very real possibility that there
is another explanation for these puzzles that does not invoke such
a surprising new element.

\vskip 3em

\centerline{{\bf Acknowledgments}}
These ideas were developed during a sequence of discussions with
a large number of people.   I am grateful to them all for their insights.

Specifically, I am happy to thank:
Tom Banks for first stressing to me that this short length scale could
be real and related to the neutrality of perturbative type II states,
and for suggesting locking; Misha Bershadsky for explaining his
ideas on conifolds; Mike Douglas for discussions of finite
momentum behavior; Dan Friedan for discussions of finite momentum
behavior and conformal field theory signatures; Jeff Harvey for
discussions about locking, finiteness
and second quantized mirror symmetry; Emil Martinec for discussions
about locking and $\CN=2$; Greg Moore for discussions about locking
and $\CN=2$; Hirosi Ooguri for describing his ideas about
conifolds;
Ronen Plesser for a discussion on compactification and logarithms;
Nati Seiberg for discussions on finiteness and light solitons;
Ashoke Sen for pointing out the utility of the dual heterotic
description; Andy Strominger for discussions about various aspects
of his work on black holes, conifolds, membranes, and
second quantized mirror symmetry,  for suggesting
locking, for explaining the sizes of various solitons
and for discussions about possible short distance cutoffs; Lenny Susskind
for discussions about his ideas on length scales in string theory,
on form factors of RR vectors, and for pointing out the significance
of size for nonperturbative soliton loop effects; Cumrun Vafa for
explaining his understanding of conifolds, for pointing out the
utility of the dual heterotic description, and for pointing out an
error in an earlier version of the arguments in section \ftt; and
Edward Witten for discussions of string duality and
for pointing out that the large size of compactified dimensions
in Einstein frame might explain the puzzle discussed in section \ftt.

Much of this work was done at the Aspen Center for Physics during
the workshop on ``Nonperturbative Dynamics of Supersymmetric Theories,''
where a preliminary version of these ideas was presented.  I would
like to thank Mike Dine and John Schwarz for organizing an unusually
stimulating workshop.

This work was supported in part by DOE grant \#DE-FG05-90ER40559.

\listrefs

\end